\documentclass{emulateapj}

\newcommand{\lta}{\lesssim}
\newcommand{\gta}{\gtrsim}

\newcommand{\ea}{et al.}

\newcommand{\gyr}{\>{\rm Gyr}}
\newcommand{\myr}{\>{\rm Myr}}

\newcommand{\msun}{\>{\rm M_{\odot}}}

\newcommand{\bdm}{\begin{displaymath}}
\newcommand{\edm}{\end{displaymath}}
\newcommand{\beq}{\begin{equation}}
\newcommand{\eeq}{\end{equation}}
\newcommand{\bit}{\begin{itemize}}
\newcommand{\eit}{\end{itemize}}
\newcommand{\ben}{\begin{enumerate}}
\newcommand{\een}{\end{enumerate}}
\newcommand{\bfi}{\begin{figure}[htb]}
\newcommand{\bpfi}{\begin{figure}[p]}

\shorttitle{Are Globular Clusters Remnant Nuclei?}
\shortauthors{B\"oker et al.}

\begin{document}


\title{Are Globular Clusters the Remnant Nuclei of Progenitor Disk Galaxies?}


\author{Torsten B\"oker}
\affil{European Space Agency, Dept. RSSD, Keplerlaan 1, 
2200 AG Noordwijk, Netherlands}
\email{tboeker@rssd.esa.int}







\begin{abstract}
The globular cluster system of a typical spheroidal 
galaxy makes up about 0.25\% of the total galaxy mass \citep{mcl99}. 
This is roughly the same mass fraction
as contained in the nuclear star cluster (or stellar nucleus) 
present in most nearby low-mass galaxies. Motivated by this 
``coincidence'', this Letter discusses a scenario in
which globular clusters of present-day galaxies are 
the surviving nuclei of the dwarf galaxies that - according
to the hierarchical merging paradigm of galaxy formation - constitute
the ``building blocks'' of present-day massive galaxies. 
This scenario, which was first suggested by \cite{fre93},
has become more attractive recently in the light of 
studies that demonstrate a complex star formation history in a 
number of massive globular clusters.
\end{abstract}




\section{Introduction}
The origin of globular cluster (GC) systems of present-day galaxies 
continues to provide a rich topic for investigation, speculation, and 
debate. Enabled mostly by the spatial resolution and precision photometry 
of the Hubble Space Telescope (HST), recent studies have established that
GCs are not uniform in their properties, but display 
a rather wide range in color and metallicity. The color distribution 
of GCs in early-type galaxies nearly always shows a 
double-peaked structure \citep{pen06}, motivating a commonly
accepted view of GC systems as ``bimodal''. The situation 
in disk-dominated galaxies is less clear, partly because the analysis
is limited by the smaller number of GCs in less massive galaxies.
While both the Milky Way and M31 show clear signs of a double-peaked
color distribution, the GC populations in more distant spirals are often
divided into ``blue'' and ``red'' subpopulations {\it by definition}
through color selection criteria \citep[e.g.][]{gou03,rho07}. 
Whether or not ``bimodality'' is universal, the broad range in GC
color makes it clear that GCs within a given galaxy
have formed neither coevally nor under the same physical 
conditions. 

Moreover, the ongoing formation of young massive 
cluster (YMCs) observed in nearby starbursts such as 
M\,82 \citep{mcc07} or NGC\,4038/9 \citep{men02}, 
as well as the presence of some intermediate-age GCs
in, e.g., the Magellanic Clouds \citep{gou06} or 
M\,33 \citep{cha06}, suggest
that GC formation was not confined to the early universe,
and possibly continues until the present day. 
Although it seems plausible
that individual YMCs can evolve into GC-like structures, 
it is less clear that entire GC systems can be created in this way.
It may well be that nature has more than 
one method to make GCs.

The recent, comprehensive review of \cite{bro06} summarizes the 
proposed scenarios for GC formation. The most widely used are
i) {\it in situ} formation through multi-phase dissipational 
collapse \citep{for97}, ii) massive star formation triggered by 
major disk-disk mergers \citep{ash92}, and iii) dissipationless 
accretion of GC systems during the assembly of massive galaxies 
through mergers at high redshift \citep{cot98}. 

These models all have considerable problems in fully explaining the
observed properties of GCs. For example, the multi-phase dissipational 
collapse model invokes two distinct phases of GC formation to explain
color bimodality, but cannot readily provide a physical mechanism that
would explain the dormant period between them. The major merger 
model, on the other hand, fails to account for the lack of 
intermediate-age GC systems and the 
systematically higher specific frequency in gE galaxies which are presumed 
to be the end products of major disk-disk mergers. Finally, the 
dissipationless accretion scenario \citep{cot98}, as well as more
recent studies employing semi-analytic models of galaxy formation
\citep[e.g.][]{bea02,rho05}, appear able to explain the observed 
color distributions of GC systems. However, these models 
do not explain how the GCs of the first galaxies formed, and 
thus merely shift the problem to earlier times.

Most importantly, none of these scenarios can explain
the observed complexity in the star formation history of
some GCs. While most ``typical''
GCs studied so far show no obvious signs of multiple stellar 
populations \citep[e.g.][]{sar07}, the situation is different for
the most massive GCs. For example, the high mass and multiple 
stellar populations of $\omega$~Cen have long triggered speculation
that it is the remnant nucleus of a satellite galaxy which was 
destroyed when merging with the Milky Way \citep{nor97,lee99}. 
Moreover, recent observations have established that $\omega$~Cen is 
not a unique case, and that the high-mass end of the Milky 
Way GC system commonly shows evidence for 
multiple stellar populations \citep{moe06,pio07,cal07}.
These results have challenged the traditional view that all
GCs are simple stellar populations born in a single,
short-lived (``instantaneous'') burst of star formation. 

This Letter revisits and expands on an alternative 
formation scenario for GCs which was first suggested by \cite{fre93}.
The underlying idea is that GCs form as the nuclei of dwarf galaxies 
in the early universe, and are accreted over time as their hosts
merge into larger structures. 
The paper is structured as follows: 
\S~\ref{sec:gc} briefly describes the \cite{fre93} scenario,
and mentions some recent observational developments that are 
relevant in this context. \S~\ref{sec:nc} reviews what is known 
about the formation of NCs in late-type disk galaxies. 
\S~\ref{sec:scenario} then discusses some quantitative aspects 
of the proposed scenario and presents a first order plausibility check.
Finally, \S~\ref{sec:issues} concludes with some open issues and predictions.

\section{Globular Clusters as Dwarf Nuclei}\label{sec:gc}
Based on the observed lack of an abundance gradient in the GC system
of the Milky Way outer halo, \cite{sea78} suggested that GCs originate 
in small ``fragments'' (i.e. proto-galaxies) which are 
accreted over an extended period of time into a larger host. 
However, they did not discuss the specifics of where and how
the GC-precursors form inside the fragments.

Building on this scenario, \cite{fre93} suggested that nucleated 
dwarf galaxies and their compact stellar cores might be the present-day 
analogs of those Searle-Zinn fragments that escaped disruption in a merger 
event. He also pointed out that chemical inhomogeneities in 
present-day GCs might be explained by self-enrichment due to the strong 
stellar winds observed in starbursting galaxy nuclei.
The general idea behind this scenario has recently received increased 
attention because it might provide a natural explanation for the observed 
multiple stellar populations in some GCs, 
especially when taking into account two other recent observational results:

1) stellar nucleation 
is not limited to dE galaxies, but is also common in low-mass disk galaxies, 
as shown by recent surveys with HST \citep{car97,boe02}. 
These nuclear star clusters (NCs) are structurally very similar to GCs,
and nearly always contain multiple stellar populations. They are therefore 
attractive candidates for the precursors of 
at least those GCs that have complex star formation histories. 
In fact, it is difficult to explain how a stellar cluster can accrete
(or retain) enough gas for multiple star formation episodes if it is not 
located at the bottom of a potential well such as the nucleus of a galaxy.

2) the mass of a GC system appears to be proportional to that of
its host galaxy \citep{mcl99}. As discussed further in 
\S~\ref{sec:scenario}, this is expected if GCs indeed form as 
the stellar nuclei of ``building blocks'' which merge into massive
present-day galaxies.

The \cite{fre93} scenario can thus be modified as follows: 
present-day GCs have formed as NCs in the centers of low-mass disk 
galaxies in the early universe where 
they grew in mass through a series of discrete star formation events
as long as the disk structure of their host remained undisturbed by
any merger activity. This growth process ends when the host galaxy 
disk is disrupted in a merger or tidal interaction. Due to its large 
stellar density, the (former) NC survives the merger, but because it 
is no longer located in a potential well, it experiences no further 
gas inflow. No further star formation takes place in the NC, and it will 
simply age passively as a GC in the merger product. 
It should be noted explicitly that the presence of multiple stellar 
populations is not a necessary consequence of this process: if
a NC was accreted early enough, it might have experienced only a single
star formation episode and thus contain only a single generation
of stars.

\section{Nuclear Star Clusters}\label{sec:nc}
Over the last decade, a number of studies - both via imaging and
spectroscopic observations - have contributed to the following 
picture of {\it present-day} NCs:

1) NCs are common: the fraction of galaxies with an unambiguous 
NC detection is 75\% in late-type (Scd-Sm) spirals \citep{boe02}, 58\% in 
earlier-type (S0-Sbc) spirals \citep*{bal07}, and 70\% in spheroidal (E \& S0) 
galaxies \citep{cot06}\footnote{All these numbers are likely lower limits, 
albeit 
for different reasons. In the latest-type disks, it is sometimes not trivial 
to locate the galaxy center unambiguously so that no particular source can 
be identified with it. In contrast, many early-type galaxies have very steep 
surface brightness profiles (SBPs) that make it difficult to detect
even luminous clusters against this bright background.}.


2) The structural properties of NCs {\it in late-type disks} 
are indistinguishable 
from those of GCs, and thus certainly do not rule out an evolutionary 
connection: typical half-light radii are $\rm 2-5\,pc$ \citep*{boe04}.
The size distribution of NCs {\it in spheroidal galaxies} also shows a large 
overlap with that of Milky Way GCs, although some present-day (d)E nuclei 
are clearly larger than the ``typical'' GC \citep{geh02,cot06}. 

3) Despite their compactness, present-day NCs are rather massive: their 
typical dynamical mass is $10^6 - 10^7\msun$ \citep{wal05} which is at the 
high end of the GC mass function. Because of their very high stellar 
densities, there is little doubt that NCs are difficult to
disrupt, and that they will remain structurally intact
even after the disruption of their host galaxy.

4) The star formation history of NCs {\it in disk galaxies}
is complex, as evidenced by the fact
that they have stellar populations comprised of multiple generations 
of stars \citep{wal06,ros06}. In the NCs of late-type disks, the youngest 
stellar generation is nearly always younger than $100\myr$ which 
explains why present-day NCs are much more luminous than 
GCs \citep{boe02,cot06}. Nevertheless, NCs always contain an old ($\sim$ Gyrs) 
stellar population, suggesting that they have been in place for a long time.

Taken together, these facts suggest that NCs have built up the bulk of 
their present-day mass through a series of star formation episodes 
\citep{wal06}. 
This repetitive ``rejuvenation'' can only occur as long as
gas is funneled towards the nucleus, and thus no longer takes place
after the NC has been removed from the galaxy nucleus or after the
gas supply has otherwise been disrupted, e.g. through transformation
of a late-type disk into a dE galaxy. 

Three recent and independent studies of NCs in different
galaxy types \citep{ros06,weh06,cot06} have established that NCs obey 
similar scaling relationships with host galaxy properties as do 
supermassive black holes. 
This has led \cite{fer06} to suggest that NCs and black holes might
constitute two different incarnations of a ``compact massive object'', 
the formation of which is inherent to galaxy assembly, and which 
contains $\sim$ 0.2\% of the host galaxy (bulge) mass. Although
\cite{bal07} have questioned the simple linearity, they confirm that  
any NC typically contains $0.1\%-0.3\%$ of the bulge mass.

A possible theoretical explanation for a similar proportionality 
also in truly bulgeless disk galaxies has 
recently been offered by \cite{ems07} who point out the compressive 
nature of tidal forces in a shallow gravitational potential, e.g. in 
an exponential disk. Based on their simulations, a NC formed
through this mechanism should contain $\sim 0.1 - 0.5$\% of the host 
galaxy mass. In this model, matter infall onto the nucleus occurs 
naturally in bulge-less disks, possibly explaining why NCs with 
recent star formation have only been found in in late-type spirals.
The following section investigates the possibility that many, if
not all, GCs are the surviving nuclei of dwarf galaxies that were 
``swallowed'' during the assembly of present-day galaxies. 
\section{A Simplistic Scenario}\label{sec:scenario}
A prerequisite for the proposed scenario is that NCs are present
already in the first galaxies, i.e. the ``building blocks'' for
galaxy assembly. This is not an implausible proposition. For example,
\cite{cen01} has suggested that the reionization of the universe produces 
inward, convergent shocks in dark matter ``mini-halos''. These shocks 
compress the baryonic gas which in turn becomes self-gravitating and 
undergoes star formation, forming a dense stellar cluster in the process. 
These ``seed clusters'' naturally define a center of gravity for the
assembly of a gas disk. As \cite{cen01} points out, the NCs 
of high-redshift mini-halos formed through this process are plausible 
progenitors of present-day GCs \citep[see also][]{moo06}.

The subsequent evolution of these ``primordial'' NCs in their parent
halos and the expected tidal fields has been studied in
detail by \cite{mas05a,mas05b}. These authors made the {\it a priori}
assumption that 0.88\% of the total halo mass is concentrated in
a NC, a somewhat higher fraction than what has been found for present-day 
NCs. These authors conclude, in agreement with \cite{cen01}, 
that the predicted properties of such NCs match those of present-day GCs, 
independent of the exact value of the initial NC mass fraction. 

They also point out that current observations are insufficient to rule out
significant amounts of dark matter in GCs which might be expected 
if GCs indeed form at the centers of mini-halos. So far, however, 
dark matter has not been convincingly detected in GCs, 
although the flat velocity dispersion profile of some
GCs found by \cite{sca07} appears consistent with the presence
of dark matter (or, as those authors prefer, with 
modified Newtonian dynamics). In any case,
even the apparent absence of dark matter in GCs doesn't fully
invalidate the idea, because \cite{mas05b} show that, depending on 
the density profile of the halo, dark matter can be
lost almost completely by tidal stripping. 

In this paper, the \cite{cen01} scenario is modified and extended by 
allowing for the 
expected secular evolution of NCs that occurs as long as they remain in 
the center of their parent halo. This secular evolution offers a natural 
explanation for the diversity in stellar populations within and across GCs.
For a first-order assessement, the following simplifying 
assumptions are made:

a) all building block galaxies are identical in that their
baryonic matter is organized in a gas-rich disk of $\sim 10^8 - 10^9\msun$.
This mass range is plausible because it makes dynamical friction an 
efficient mechanism to merge small galaxies into a larger halo \citep[see Eq. 7-27
in][]{bin87}. It is also consistent with 
models of structure formation \citep[e.g.][]{moo06}, as well as with observed
masses of dwarf galaxies in the local universe \citep[e.g.][]{geh06}.

b) the center of each progenitor disk is occupied by a NC with $\sim$0.2\% 
of the host mass. As long as the disk escapes merger activity, its NC
grows in mass via a series of nuclear starbursts fueled by gas flow 
from the disk onto the nucleus, similar to those observed in nearby 
galaxies \citep[e.g.][]{sch03,sch06,sch07}.

c) once the disk experiences a destructive merger, the NC is captured in 
the halo of the merger product. From then on, it ages passively
as a GC. Therefore, the later a galaxy is accreted, the more massive the 
newly added GC, and the more complex its star formation history.

d) stars and gas of the merging disk will be incorporated 
into the merger products' halo and disk, respectively, so that
the total mass of the merger product is simply the sum of all
the constituents.

e) the potential for cluster formation induced by the merging process
is ignored for the moment, i.e. GCs are {\it only} added through 
accretion of NCs.

In this simplistic scenario, the total number of GCs is equal
to the number of accreted building blocks. Because any
intermediate merger products are created from identical building blocks
as described in assumption a), the number of GCs simply scales 
linearly with assembled galaxy mass, regardless of the
details of the merger tree. This offers a natural explanation for
the empirical result that all but the latest-type spirals 
(see \S\ref{sec:issues}) show a 
constant specific frequency of GCs: the observed T-value \citep{zep93} is
$\sim$ 2 GCs per $10^9\msun$ \citep{gou03,rho07}. This value is likely
a lower limit because i) not all GCs will survive and ii) the faintest 
GCs might be missed in surveys. Taken at face value, however, it follows 
that the building blocks should have masses of 
$\lta 5\times 10^8\msun$, entirely consistent with assumption a) above.

The primordial NC of such a galaxy should then contain $0.002\times
5\cdot 10^8\msun$, i.e. $\lta 10^6\msun$, in good agreement with the 
smallest observed NC mass of $8\times 10^5\msun$ \citep{wal05}. This 
apparent minimum NC mass is unlikely to be a selection effect, because
\cite{boe02} found a faint-end cutoff in the NC luminosity function 
in a much larger, unbiased sample of NCs. 
Note that NC masses in this range lie well
above the critical mass $m_{*}\sim 2\times 10^5\msun$ at which 
the GC mass function changes slope because less massive clusters
are prone to disruption \citep{fal01}. It is therefore reasonable to 
expect that any such clusters will have survived until the present day.

Nevertheless, depending on the details of their orbit and initial density 
profile, the captured clusters will experience two-body relaxation and/or 
tidal stripping. These effects can easily reduce the cluster mass by 
a factor of 2-5 over $\sim 10\gyr$ \citep*[e.g.][]{ves97}. 
An even higher mass loss has been reported in some Milky Way 
GCs \citep{dem06,dem07}, making dynamical evolution a likely explanation
for the fact that many GCs today have masses below $10^5\msun$ \citep{mcl05}.

If present-day GC systems are indeed made from NCs
accumulated throughout the merging history of a galaxy, then the mass 
of any GC system should - to first order - contain the same mass fraction 
as each individual NC (because mass is conserved according to 
assumption c) above), i.e. $\sim$0.2\% of the total host galaxy mass. 
Intriguingly, this appears entirely consistent with observations: 
\cite{mcl99} concludes that the formation efficiency of 
GCs is universal, at least for ellipticals and the spheroidal
component of disk galaxies (incl. the halo of the Milky Way), with a value of 
$\epsilon_{\rm cl}\equiv M_{\rm GCs}/(M_{\rm gas}+M_{\rm stars})\sim 0.0026$.
It should be noted, however, that the situation is less clear when entire 
disk galaxies (i.e. including the gas-rich thin disk) are considered \citep{bro06}. 

Assuming for the moment that $\epsilon_{cl}$ is indeed 
universal, and had the same value at all redshifts, 
the argument can be turned around: 
the putative GC system of a $10^8\msun$ building block galaxy should 
then contain a mass of only $5\times 10^5\msun$. Unless this mass is 
concentrated in very few (one or two) clusters, none of the GCs in that
system will have survived until the present day. Note that this
also represents a challenge for the dissipationless accretion scenario, 
because in order to produce significant numbers of stellar clusters 
{\it in the halos} of early-universe galaxies with enough mass to 
survive until z=0, the value of $\epsilon_{cl}$ must have been 
significantly higher in the past.


\section{Open Issues and Predictions}\label{sec:issues}
The toy model described in the last section is clearly
simplistic, and there are a few points that require further investigation.
For example, it has been assumed throughout the paper that the mass
fraction of NCs remains constant over time at 0.2\% of the host galaxy, 
as observed {\it in present-day galaxies}. For this assumption to hold,
the gas accretion rate onto the host galaxy must match 
the mass growth of the NC. To first order, such a link does not appear 
implausible because there are a number of mechanisms that transport gas
from the outskirts of a galaxy towards the center. However, it is presently 
unclear whether galaxy evolution models can indeed explain such 
``fine tuning''.

Moreover, there is no doubt that the stellar populations of GCs are 
generally old. For example, the Milky Way GCs have ages of $\gta 8\gyr$ 
with a relatively small spread of $< 3\gyr$ \citep{dea05}. Because in the
proposed scenario, the age of {\it the youngest stellar population} in
a GC denotes the time when its former host was disrupted, this implies 
that the merging of disk galaxies into the Milky Way was mostly 
complete $8\gyr$ ago. Any NCs accreted after that time must have been 
residing in dE-type galaxies, where they had already been aging passively 
for some time because there is no gas supplied to the nucleus in these
galaxies. While this conclusion does not seem to be contradicted 
by any current observations, more detailed simulations of the Milky Way
formation are required to test its plausibility.

A third issue that is not explained by the proposed scenario
is the double-peaked color distribution of many GC systems.
In massive galaxies, where this effect is most pronounced, it may
be possible to invoke additional cluster formation
during the merger of two massive spirals (each of which has formed its
own GC system according to the proposed scenario), but {\it not} in any 
minor mergers that led to the build-up of the disk galaxies themselves.
That star formation is indeed enhanced in major mergers is
undisputed and has been observed in a number of nearby objects 
\citep[e.g.][]{whi93,whi95}. However, the fact that there are hardly any
GCs known with intermediate ages (few Gyr), requires that the
majority of these mergers were completed by $z\sim 2$, possibly a 
challenge to models of galaxy formation. On the other hand, it
may not even be necessary to invoke the presence of two
GC sub-populations with intrinsically different metallicities 
to explain color bimodality \citep*{yoo06}.

One prediction that the proposed scenario can be tested
against is that ``pure'' disk galaxies should have a 
systematically lower T-value. This is because they
apparently had an uneventful merger history which enabled them to
avoid the formation of a bulge. Because these galaxies likely assembled
a larger fraction of their mass through slow accretion of intergalactic 
gas rather than via accretion of satellites, they should have gathered 
proportionally fewer NCs. Generally speaking, little is known about the GC 
systems of the latest-type spirals, although a few studies have been
undertaken which seem to confirm this prediction. 
For example, \cite{van82} report the absence of GCs in the edge-on 
spiral NGC\,891 which, however, might not be a good example of a
bulge-less system since it is classified as Sb. 
Moreover, \cite{ols04} find only a 
handful of confirmed GCs in most late-type spirals of the Sculptor group,
and report that the specific frequency of GCs is lower in late-type
spirals than in bulge-dominated galaxies. Unfortunately, these authors 
do not present T-values which is a more robust diagnostic because of 
potential color variations between galaxy types.

It is still uncertain when exactly
NCs formed in the universe and what the duty cycle of rejuvenation
in those early nuclei was. In principle, one would expect that
a large fraction of GCs show some level of complexity in their 
stellar populations.
However, the earlier a NC was acquired, the fewer stellar generations 
it should contain, and the smaller the expected age differences between 
those populations are. As mentioned in \S~\ref{sec:gc}, single-population 
GCs are certainly expected, especially if there
was a considerable delay between the original collapse of the
\cite{cen01} model, and the onset of ``rejuvenation'' of the NC.


On average, however, more massive GCs should have been accreted more 
recently than low-mass ones.
Because they spent a longer time in a galaxy nucleus, they experienced
a higher number of star formation events, and contain both younger and 
more complex stellar populations. The fact that multiple stellar 
populations so far have mostly been detected in rather massive GCs 
such as NGC\,2808 and $\omega$~Cen thus is not surprising.

In summary, the speculative scenario for GC formation laid out in this 
letter does not seem implausible, and offers a natural explanation 
for the observed mass of GC systems 
($M_{\rm GCs}\approx 0.0025\cdot M_{\rm gal}$)
and the presence of multiple stellar populations in massive GCs.



\acknowledgments
I am grateful to C. J. Walcher, R. van der Marel, G. de Marchi,
and M. Perryman for helpful comments. The anonymous referee is 
acknowledged for a thorough and critical review, as well as a number
of detailed suggestions which greatly improved this paper.

\end{document}